# Permittivity-asymmetric qBIC metasurfaces for refractive index sensing

*Xingye Yang[1], Alexander Antonov[1], Haiyang Hu[1*], Andreas Tittl[1*]*

1. Chair in Hybrid Nanosystems, Nanoinstitute Munich, Faculty of Physics, Ludwig-Maximilians-Universität München, Königinstraße 10, 80539 München, Germany.

*E-mail: Andreas.Tittl@physik.uni-muenchen.de
*E-mail: Haiyang.Hu@physik.uni-muenchen.de

## Abstract

Bound states in the continuum (BICs) provide exceptional light confinement due to their inherent decoupling from radiative channels. Small symmetry breaking transforms BIC into quasi-BIC (qBIC) that couples to free-space radiation enabling ultra-high-quality-factor (Q-factor) resonances desirable for refractive index (RI) sensing. In practical implementations, geometric asymmetry is typically employed. However, since the radiative loss remains fixed once fabricated, such metasurfaces exhibit only a horizontal translation of the resonance spectrum in RI sensing, without modification of its overall shape. Here, we demonstrate a permittivity-asymmetric qBIC (ε-qBIC) metasurface, which encodes environmental refractive index variations directly into the asymmetry factor, resulting in indexes response involving both resonance wavelength shift and modulation variation. In addition to exhibiting a competitive transmittance sensitivity of ~5300%/RIU under single-wavelength conditions, the ε-qBIC design provides a substantially improved linear response. Specifically, the linear window area of its sensing data distribution, calculated as the integrated wavelength region where the linearity parameter remains above the preset threshold, is 104 times larger than that of the geometry-asymmetric qBIC (g-qBIC), enabling more robust and reliable single-wavelength signal readout. Additionally, numerical results reveal that environmental permittivity asymmetry can optically restore the g-qBIC to a state with ultra-high-Q (over $10^7$), approaching to BIC condition. Unlike traditional BICs, which are typically inaccessible once perturbed, the permittivity-restored BIC becomes accessible through environmental perturbations. These findings suggest an alternative design strategy for developing high-performance photonic devices for practical sensing applications.

## 1. Introduction

Among the various optical resonances in nanophotonics, bound states in the continuum

(BICs) stand out for their ability to confine light without radiative loss, enabled by symmetry or interference that decouples them from free-space modes [1], [2]. In particular, if the coupling constants vanish due to symmetry, such BIC are symmetry-protected BICs [3], [4], [5]. To make these idealized, non-radiative states accessible in practical devices, a slight symmetry-breaking is typically introduced, transforming them into quasi-BICs (qBIC) that retain high quality factor (Q factor), which is suitable for compact devices [6], [7], [8], [9]. Conventional implementations usually achieve this by introducing minor geometrical asymmetries, such as tilted elliptical rods[10], nanodisks with off-centered holes[11], or asymmetric split-rings [12] , among other structures [13], [14], [15].

Among the many emerging applications, refractive index (RI) sensing based on symmetry-protected BIC metasurfaces has recently attracted growing interest. Compared to other types of metasurfaces[16], [17], [18], [19], qBIC resonances offer the advantage that their Q factor can be tuned simply by adjusting the geometrical asymmetry [20], [21], [22], [23], [24]. However, like most resonance-based sensing platform, the readout method for the environmental RI changes relies on tracking the spectral shift of an optical resonance due to changes in the surrounding medium [25], [26]. It typically requires high-resolution spectrometers and stable broadband light sources, which add cost and complexity [27], [28].

To address these practical constraints, recent efforts have explored single-wavelength intensity variation, where the sensing signal is extracted from the intensity modulation at a fixed probe wavelength near the resonance [27], [29], [30]. Such an approach simplifies the hardware and enables integration into compact, low-cost platforms.

However, for conventional geometry-asymmetric qBIC (g-qBIC) metasurfaces, the radiative loss is essentially fixed during sensing. In low-loss environments, changes in the surrounding refractive index do not alter the radiative loss, and thus mainly induce a lateral spectral shift with minimal impact on the vertical modulation depth of the resonance. [31], [32], [33], [34], [35].

In contrast, permittivity asymmetric qBIC metasurfaces can offer a fundamentally different mechanism for interacting with RI environment. It is well know that radiative coupling of qBICs with far-fields is governed by asymmetry factor, which conventionally defined by structure's geometrical asymmetry as mentioned before [21], [24]. However, according to our previous research, based on the permittivity asymmetric qBIC (ε-qBIC) metasurfaces, we can encode the environmental RI into asymmetry factor of the system [36]. As a result, the qBIC becomes highly responsive to RI variations, resulting not only a shift of resonance position, but also a pronounced modulation of its amplitude in the low loss system. This response improves the linearity of the sensing signal in single-wavelength analysis, which, while not strictly required for all sensing approaches, facilitates simpler and more robust signal interpretation compared to nonlinear sensing data distributions with environmental RI, thereby enhancing sensing stability.

In this work, the sensing performance of ε-qBIC metasurfaces is demonstrated, and the concept of an environment-accessible restored BIC enabled by permittivity asymmetry is proposed through simulations. Simulations and experiments first confirm that the ε-qBIC metasurface exhibits not only resonance wavelength shifts but also additional resonance modulation with changes in the RI, in contrast to the purely wavelength-shift response of g-qBIC's spectrum, using commercially available refractive index oils ($\Delta n = 0.02$). To evaluate the relative quality of signals between wavelength shift and single wavelength intensity modulation, sensing experiments were further conducted with smaller RI intervals ($\Delta n = 0.004$). The intensity variation signal ($\Delta T_s$) at a probe wavelength achieved a higher signal-to-noise ratio (SNR ≈ 17 dB) compared to the wavelength-shift signal ($\Delta \lambda_s$) (SNR ≈ 5 dB).

Consequently, the intensity variation signal ($\Delta T_s$) was further analyzed, and the experimental responses of ε-qBIC and g-qBIC metasurfaces were compared. Comparable sensitivities were obtained for both metasurfaces (~5000%/RIU). However, the ε-qBIC metasurface exhibited a markedly improved linearity of the sensing response, with a linear window, defined as the wavelength range where the extracted linearity parameter exceeds the preset threshold, whose integrated area is approximately 104 times larger than that of the g-qBIC. This wider window indicates lower noise across a broader wavelength range, thereby improving robustness and stability for practical sensing applications.

Finally, from a fundamental perspective, numerical results reveal that geometric asymmetry, which typically converts BIC states into quasi-BICs, can be compensated by precisely tailoring the environmental permittivity profile within the unit cell. This enables the recovery of BICs with radiative losses approaching zero. Notably, the restored symmetry protected BIC states (RSP-BIC[37]) become optically accessible to changes in the surrounding environment, an interaction that is otherwise forbidden for conventional BICs. These findings suggest a broader design strategy: radiative channels in BIC systems can be tuned not only through geometry but also via environmental permittivity engineering, enabling new possibilities in high performance photonics platform for practical sensing application.

## 2. Results and discussion

### 2.1 Environmental RI-controlled asymmetry factor in ε-qBIC metasurfaces

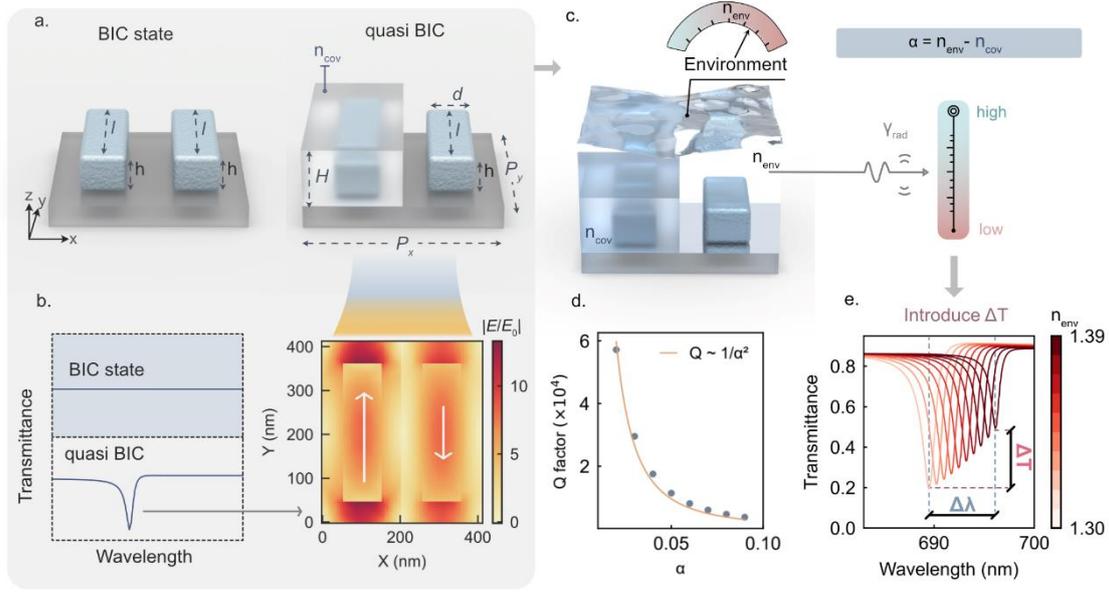

**Figure 1**. **The concept of environmental permittivity-asymmetric quasi-BIC metasurfaces for refractive index sensing. a**. Illustrating the unit cell of the ε-*q*BIC metasurfaces, consisting of two identical dielectric nanorods made of $TiO_2$ with periodicities $P_x = P_y = 410$ nm. The length *l*, width *d*, and height h of the nanobar are 320 nm, 105 nm, and 110 nm respectively. One of the resonators is encapsulated in a dielectric medium with refractive index of $n_{cov}$ and the height of *H*. **b**. Transition from BIC to ε-qBIC by adding a cladding layer on one of the nanobars, and corresponding electric field distribution of the ε-qBIC. **c**. Asymmetry factor defined by $α = n_{env} – n_{cov}$, where $n_{cov} = 1.49$ and $n_{env}$ represent the refractive index of the dielectric medium covering on the one of nanorods, and the global environment of the whole ε-qBIC metasurfaces respectively. **d**. The Q factor exhibits an inverse-square dependence on the asymmetry factor, indicating that the asymmetry factor can modulate the radiative loss. **e**. Spectral response under varying environmental refractive index (RI), where changes in radiative loss additionally induce variations in transmittance intensity.

To construct the ε-qBIC resonance mode in our metasurface, the design follows two steps. First, the unit cell is composed of two identical $TiO_2$ nanorods, which support a symmetry-protected BIC due to their structural symmetry. Second, a dielectric cover layer is selectively applied to one of the rods, introducing a difference in permittivity between the surroundings of the nanorods. This breaks the symmetry in the permittivity distribution and transforms the original BIC into a quasi-BIC, accompanied by a radiative resonance and a distinct electric field profile (Figure 1a–b).

The asymmetry in permittivity leads to unequal dipole strengths between the two out-of-phase electric dipoles, resulting in a nonzero net dipole moment. This enables the resonance to couple to the far field and become observable in the transmission spectrum. Specifically, the two $TiO_2$ nanorods have dimensions of $l = 320$ nm in length, $d = 105$ nm in width, and h = 110 nm in height. One of the rods is fully encapsulated in a dielectric medium with refractive index $n_{cov}$, while the other remains exposed to an external medium with refractive index $n_{env}$. For example, this can be implemented using a PMMA layer ($n_{cov} = 1.49$) with height *H* of 300 nm as the cladding material.

The prepared metasurface is then placed in the sensing environment (Figure 1c). Unlike conventional qBIC designs, such as those involving tilted elliptical rods or rods with different lengths, where the structural asymmetry, and thus the radiative loss, is fixed after fabrication, the ε-qBIC structure introduces asymmetry through the refractive index contrast between the cladding and the environment. This contrast is quantified by the asymmetry factor $\alpha = n_{env} - n_{cov}$, allowing the radiative loss to vary dynamically with the surrounding medium. This work first demonstrate in RI sensing where environmental changes are directly translated into both the asymmetry factor and the associated radiative loss. As shown in Figure 1d, the Q-factor exhibits an inverse-square dependence on the asymmetry factor (related with environment RI), consistent with the characteristic behavior of qBIC modes. This dependence is also reflected in the optical spectra: as the refractive index of the environment varies (from 1.30 to 1.39), the system exhibits not only a shift in resonance wavelength but also a modulation in resonance strength. (Figure 1e). This feature improves the linearity of the sensing signal readout, resulting in better stability over a broader wavelength range. We will return to this point later. Before that, we experimentally verify the simulation results, as discussed in the next section.

## 2.2 Experimental validation of both resonance wavelength and intensity modulation in ε-qBIC metasurface

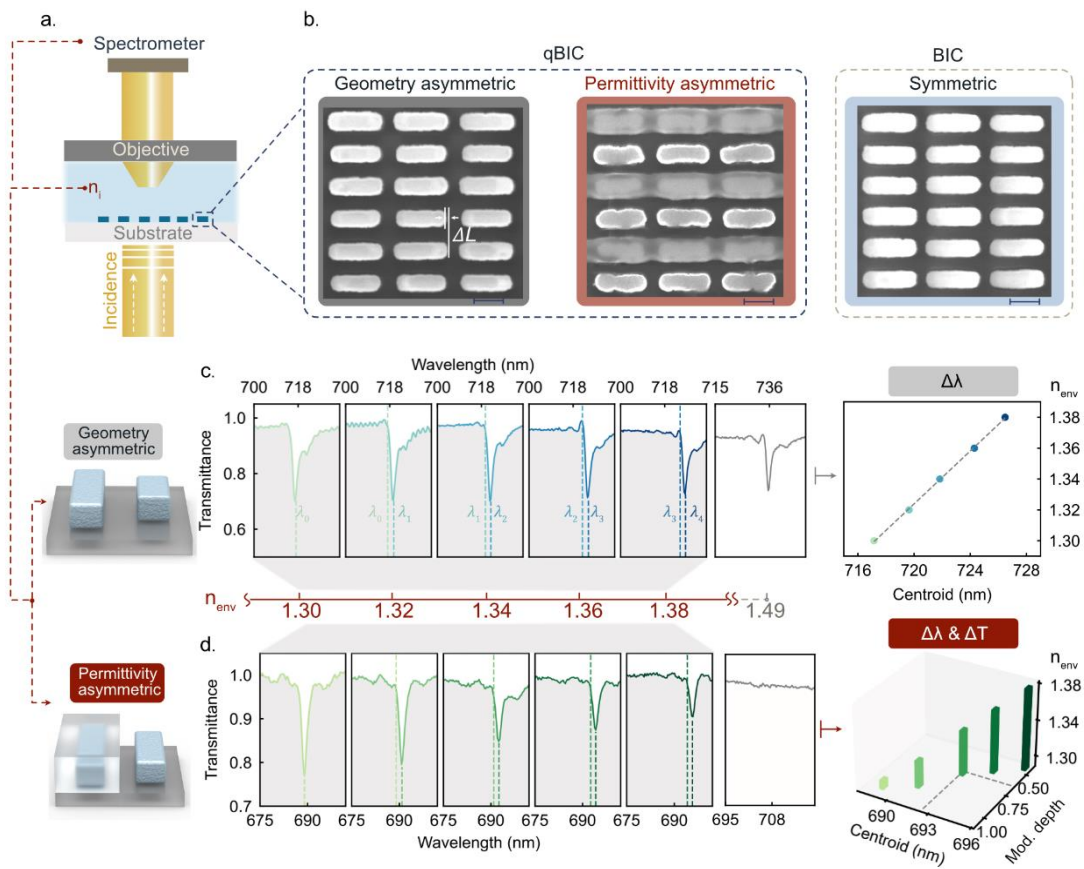

**Figure 2. Experimental characterization of the qBIC metasurfaces for refractive index sensing. a**.

Schematic of the experimental setup for transmittance measurement for g-qBIC and ε-qBIC metasurfaces in an aqueous environment. **b**. The SEM images for the g-qBIC metasurfaces, ε-qBIC metasurfaces, and BIC metasurfaces. **c**. Measured transmittance spectra of g-qBIC metasurfaces immersed in commercial index-matching oils with refractive indices $n_i$ = 1.30, 1.32, 1.34, 1.36, 1.38. The PMMA ($n_{cov}$ = 1.49) serves as the cladding material. The corresponding resonance shifts are analyzed using a centroid method, shown on the right, where $\lambda_0$ to $\lambda_4$ denote the centroid wavelengths of the transmission dips, highlighting a clear monotonic wavelength shift as the RI increases. **d**. Measured transmittance spectra of ε-qBIC metasurfaces under the same set of environmental refractive indices. In addition to the spectral shift, the modulation depth of the resonance also varies with $n_i$. The right panel presents changes in both centroid wavelength and normalized modulation depth.

To experimentally validate the refractive index sensing capabilities of both g-qBIC and ε-qBIC metasurfaces, we performed transmittance measurements (Figure 2a) in a controlled aqueous environment using commercial index-matching oils with well-defined refractive indices (details in Method). Both metasurfaces share the similar Q factor measured in the air to ensure the fair comparison (Figure S2). Their fabrication have been confirmed through the SEM images (Figure 2b). The fabrication process of ε-qBIC can be found in Figure S1.

The transmittance spectra of g-qBIC metasurfaces are shown in Figure 2c, measured across a series of index-matching oils ($n_i$ = 1.30, 1.32, 1.34, 1.36, 1.38) with a refractive index step size of $\Delta n$ = 0.02. As expected, the resonance wavelength exhibits a monotonic redshift with increasing environmental index, consistent with the behavior of conventional single-parameter high-Q RI sensing platforms. The centroid analysis provides the wavelength centroid of a resonance within a half-wavelength range, enabling a clearer tracking of its wavelength shift under different refractive indices (Method of centroid analysis in Figure S2). The centroid analysis on the right of Figure 2c confirms linear trend, with the spectra shift of around 2 nm per RI step. Notably, the resonance modulation depth remains nearly constant, even at $n_i$ = 1.49, indicating that the optical response is primarily governed by phase effects rather than radiative coupling changes.

In contrast, Figure 2d demonstrates the spectral response of ε-qBIC metasurfaces under the same set of environmental refractive indices. The resonance wavelength shows a redshift of approximately 2 nm with increasing $n_i$, similar to the g-qBIC case. However, a key difference lies in the modulation depth, which gradually decreases and eventually disappears at $n_i$ = 1.49. This critical point corresponds to a fully permittivity symmetric configuration, where the system reverts to a BIC state. The complete disappearance of the resonance provides a distinct optical signature, highlighting the strong potential sensing capability enabled by permittivity-induced asymmetry.

This variation in modulation depth reflects the dependence of the Q-factor on the asymmetry factor α, as previously discussed in Figure 1c to e. The right panel of Figure 2d provides a quantitative analysis of both the resonance centroid and the modulation depth, clearly showing that ε-qBIC metasurfaces exhibit not only a resonance wavelength shift but also a change in resonance intensity.

After validation of the simulation results, the next step is to assess two common sensing readouts: the resonance wavelength shift and the transmittance change at a fixed wavelength. The analysis focuses on determining which signal provides a stronger and more robust response under background noise for the ε-qBIC metasurface.

## 2.3 Single-wavelength intensity modulation outperforms wavelength shift in signal robustness

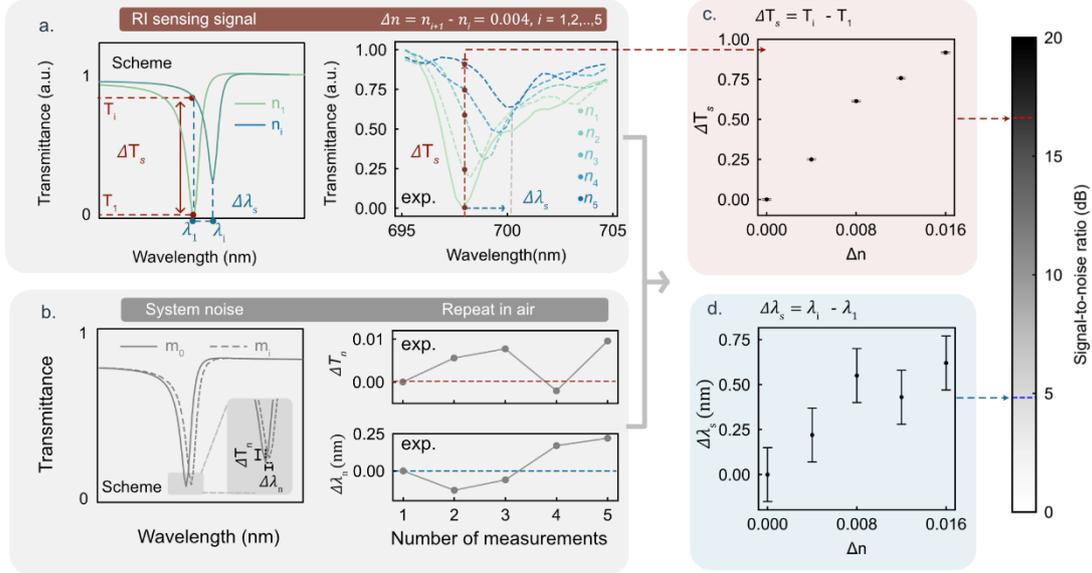

**Figure 3. Comparative analysis of signal quality between wavelength shift and intensity variation in RI fine sensing. a**. Schematic representation of two types of signals used in RI fine sensing, along with normalized experimental spectra. The intensity variation $\Delta T_s$ refers to the transmittance difference at a fixed wavelength, while $\Delta \lambda_s$ denotes the shift in resonance wavelength. Experiments were conducted in five refractive index environments ($n_1$ to $n_5$), ranging from 1.386 to 1.402, stepsize $\Delta n$ = 0.004. **b.** Illustration of system noise, showing small variations in transmittance ($\Delta T_n$) and resonance position ($\Delta \lambda_n$) based on repeated measurements ($m_0$ to $m_i$) under identical air conditions. Quantitative evaluation of noise levels was performed across five repeated measurements on the same ε-qBIC metasurface. **c–d.** Signal-to-noise ratio (SNR) analysis for $\Delta T_s$ and $\Delta \lambda_s$ based on the extracted signals in (**a**). Error bars represent the experimentally determined system noise. The signal $\Delta T_s$ exhibits a significantly higher SNR, indicating better robustness for sensing applications.

To evaluate which of the two available signal channels in the ε-qBIC metasurface, resonance wavelength shift or intensity variation at a fixed wavelength, offers better performance in RI sensing, we conducted a set of high-resolution sensing experiments, by narrowing the refractive index step, to extract the signal-to-noise ratio (SNR) under system noise conditions.
In this experiment, glycerol–water mixtures with refractive indices from 1.386 to 1.402 were prepared in increments of $\Delta n$ = 0.004. These five RI values ($n_1$ to $n_5$) correspond to the spectra presented in Figure 3a. To highlight the relative changes between spectra, the transmission spectrum from the first measurement ($n_1$) was taken as a reference. Its maximum and minimum values were normalized to 1 and 0, respectively, so that all subsequent spectra could be compared on the same scale. This normalization preserves the overall spectral shape while enabling the relative variations to be clearly observed. For each RI, the transmission spectrum was recorded, and two sensing parameters were extracted. The intensity variation signal $\Delta T_s$ is defined as the change in transmittance at a fixed probe wavelength, $\Delta T_s = T_i - T_1$, where $T_i$ is the normalized transmittance for $i$-th RI. The resonance shift signal $\Delta\lambda_s$ is defined as the change in resonance wavelength where $\lambda_i$ represent resonance wavelength extracted from the $i$-th measurement.

To evaluate the system noise, five repeated measurements in air were performed after each sensing step. Prior to each repetition, the sample was rinsed, dried and measured under identical conditions, simulating practical scenarios involving sample reuse and capturing typical instrumental and operational fluctuations. The corresponding variations in transmittance ($\Delta T_n$) and resonance wavelength ($\Delta\lambda_n$) are presented in the right panel of Figure 3b, with a conceptual illustration provided in the left. These repeated measurements were used to quantify the noise levels, which were then incorporated as error bars in the analysis of $\Delta T_s$ and $\Delta\lambda_s$. As shown in Figure 3c, the intensity variation signal exhibits consistently higher SNR values across the tested RI range, exceeding 15 dB. By contrast, the resonance shift signal $\Delta\lambda_s$ also, although above the noise threshold, displays greater fluctuations and a lower average SNR (Figure 3d). These results indicate that, when accounting for measurement repetition and associated operational fluctuations, the intensity variation channel provides a more robust signal. In the following section, the advantages of using $\Delta T_s$ for RI sensing with ε-$q$BIC metasurfaces are examined in comparison with g-$q$BIC.

## 2.4 Higher linearity of ε-qBIC sensing data distribution under single wavelength detection

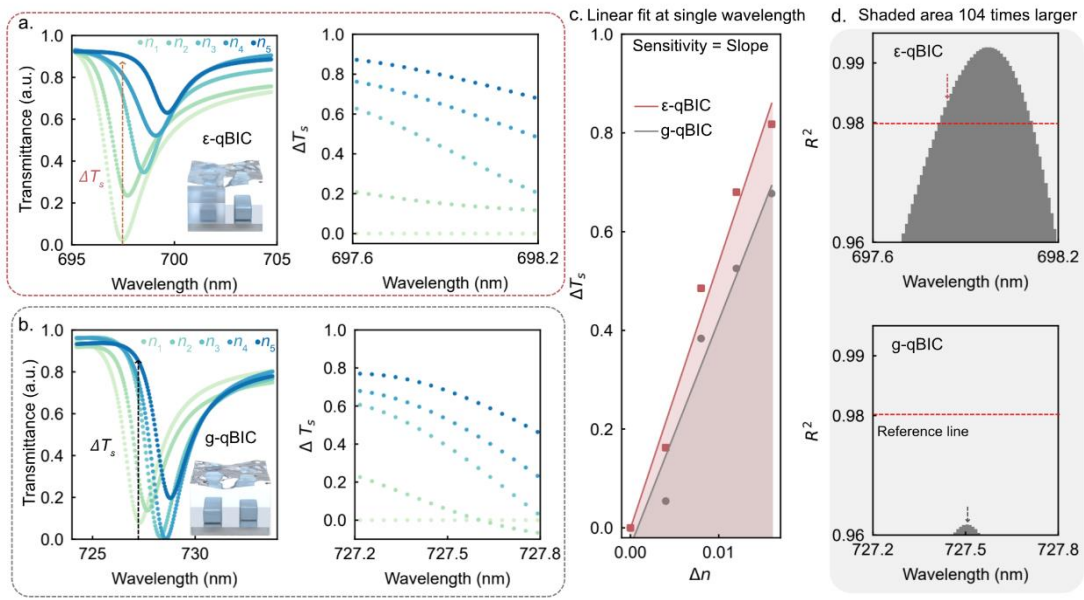

**Figure 4. Comparison of sensing performance between ε-qBIC and g-qBIC metasurfaces under

**identical RI conditions.** Refractive indices $n_1$ to $n_5$ range from 1.386 to 1.402, with a step size of $\Delta n = 0.004$. **a**. Experimental transmittance spectra of the ε-qBIC metasurface under different RI environments, along with the extracted intensity variation signals. **b**. Transmittance spectra and corresponding intensity variation signals for the g-qBIC metasurface under the same RI conditions. **c**. Linear fitting of intensity variation signals at a single representative wavelength, comparing the responses of ε-qBIC and g-qBIC. **d**. $R^2$ values obtained from linear fitting across multiple wavelengths, used to evaluate the linearity of the intensity variation response. Higher $R^2$ values indicate better linearity and lower fitting residuals. The ε-qBIC metasurface shows consistently higher $R^2$ values across a broader wavelength range, indicating enhanced linearity.

To further evaluate the sensing performance of ε-qBIC metasurfaces based on intensity variation signals, we conducted a direct comparison with g-qBIC under identical refractive index conditions ($n_1$ to $n_5$, ranging from 1.386 to 1.402 with a step size of $\Delta n = 0.004$). Figures 4a and 4b show the transmittance spectra and extracted intensity variation signals for ε-qBIC and g-qBIC, respectively.

Linear fitting of the intensity variation $\Delta T_s$ across the five RI values at each wavelength was performed. Figure 4c compares the fitted results for both metasurfaces at a selected representative wavelength. The slope of each fit corresponds to the sensitivity ($\Delta T_s / \Delta n$), while the coefficient of determination $R^2$ is used to evaluate the fitting quality. The $R^2$ value is defined as $R^2 = 1 - S_{res} / S_{tot}$, where $S_{res} = \Sigma(y_i - \hat{y}_i)^2$ is the residual sum of squares between the measured values $y_i$ and the fitted values $\hat{y}_i$, and $S_{tot} = \Sigma(y_i - \bar{y})^2$ is the total sum of squares with respect to the mean $\bar{y}$ of all $y_i$. $y_i$ is the corresponding sensing data $\Delta T_s$. A higher $R^2$ value for a fitting indicates better linearity and lower noise in the signal.

As shown in Figure 4c, both ε-qBIC and g-qBIC exhibit high sensitivity in the range of ~5000% / RIU. While some recent studies using metallic sensing structures have demonstrated extremely high spectral resolution [38], the sensitivity values reported in recent works based on dielectric metasurfaces, particularly those employing single-wavelength transmittance variation readout, are typically around 2000% / RIU [29], [39], [40].

For the g-qBIC, the maximum $R^2$ observed across the wavelength range is 0.9617, at 727.5 nm. Since none of the wavelengths for g-qBIC reach $R^2=0.98$ which is the commonly referenced benchmark [41], [42], [43], we selected the highest $R^2$ point for comparison. In contrast, ε-qBIC achieves a competitive sensitivity (~5382% / RIU) at a wavelength where $R^2 = 0.9813$, exceeding the 0.98 threshold. This indicates that ε-qBIC not only offers excellent sensitivity but also delivers improved signal quality and linearity.

Moreover, simulations in Figure S4 suggest that the sensitivity of ε-qBIC can be further enhanced by adjusting the cladding index $n_{cov}$ to closely match the refractive index of the sensing environment. In such high-Q regimes, even a small RI change can significantly alter the radiative loss channel to 0 and restore the ε-qBIC resonance to a BIC state, as the metasurface effectively exhibits symmetry both in geometry (identical rods) and in the surrounding dielectric environment. This leads to a sharp transition from a resonant to a non-resonant state. This results in a unity intensity modulation at a fixed wavelength, representing an ideal condition for signal readout in RI sensing applications.

Beyond individual points, we further compared the $R^2$ distribution across the wavelength range (Figure 4d). While g-qBIC fails to exceed threshold within the range, ε-qBIC consistently maintains $R^2$ values above it. To quantify the effective linearity window, we applied a relaxed threshold of $R^2 = 0.96$ and calculated the integrated area between the $R^2$ curve and the threshold line. The ε-qBIC shows an area approximately 104 times larger than that of the g-qBIC, clearly demonstrating its much broader and more stable sensing window with reduced sensitivity to noise.

**2.5 Environmentally permittivity restored symmetry-protected BIC**

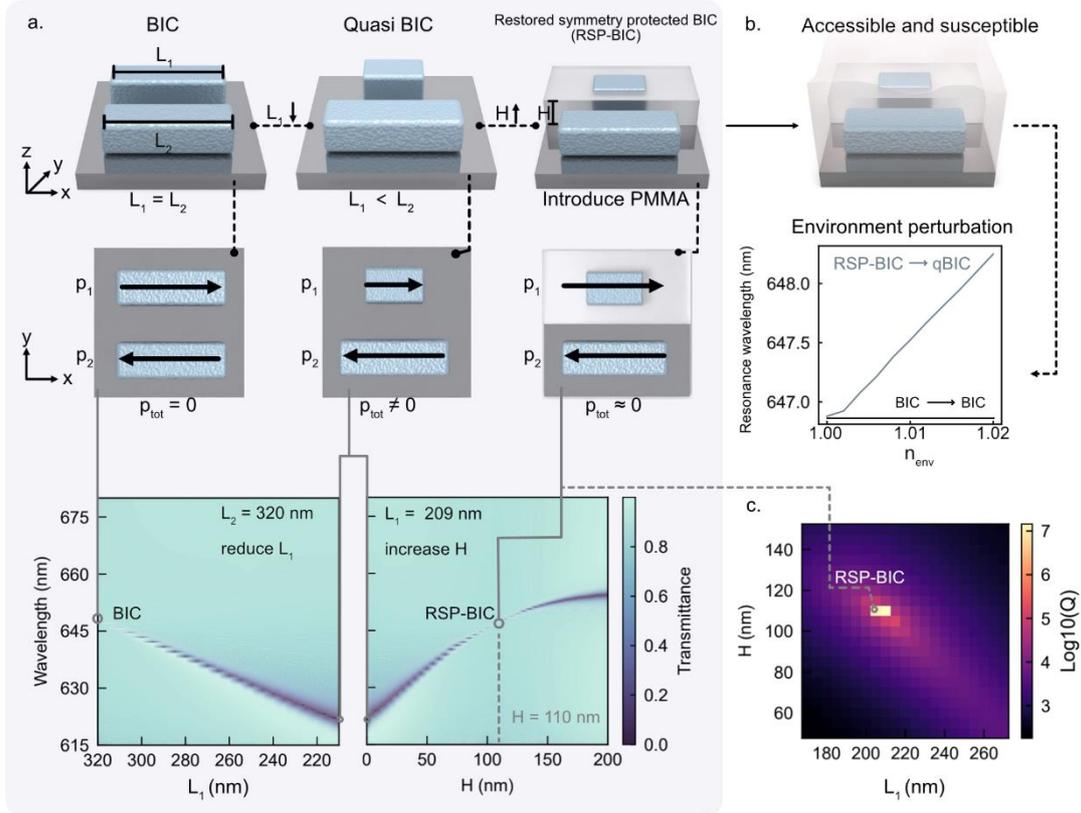

**Figure 5. Permittivity-controlled restoration of symmetry-protected BIC states. a**. Conceptual illustration of transformation of g-qBIC into restored symmetry-protected BIC (RSP-BIC) via environmental permittivity control. Starting from a symmetric structure supporting a BIC (left), introducing a geometric perturbation (shorter $L_1$) breaks the in-plane symmetry and produce a g-*q*BIC state. By selectively covering the shorter resonator with a dielectric layer of refractive index $n_{cov}$, the permittivity asymmetry compensates the geometric imbalance, restoring the BIC state (right) with radiative losses approaching to zero. The middle panel illustrates the corresponding electric dipole distributions, while the bottom panel shows the associated transmittance spectra during this process. Simulated transmittance spectra showing the evolution of a BIC under geometric perturbation ($L_1$) and its recovery via covering the shorter rod ($L_1 = 209$ nm) with growing height *H* of a dielectric layer ($n_{cov}=$ 1.5). **b**. Transforming of RSP-BIC into radiative qBIC due to immersing the metasurface into environment with $n_{env}$. In contrast, conventional BICs are optically inaccessible and remain completely

isolated from the environment with $n_{env}$. **c**. Calculated *Q* factor map as a function of rod length $L_1$ and dielectric cover layer height *H*, demonstrating the condition for RSP-BIC state characterized by a *Q* factor $>10^7$. This indicates the compensation point where the permittivity asymmetry cancels the geometric perturbation.

We demonstrate that such a g-qBIC state can be optically restored to a close to BIC state by applying a compensating permittivity asymmetry. Specifically, the simulations start from a structure with the same period, resonator height, and width as used in Figure 1. The initial configuration is a symmetric geometry with equal rod lengths ($L_1 = L_2 = 320$ nm), as shown on the right side of Figure 5a. In this case, the effective dipole moments ($p_1$ and $p_2$) cancel each other, resulting in a decoupled state with no resonance observable in the transmittance spectrum.

Next, by decreasing the length of $L_1$, we break the in-plane symmetry, resulting in a nonzero net dipole moment. This leads to a radiative qBIC mode that becomes observable as a resonance peak in the spectrum.

We then fix the short rod length ($L_1 = 209$ nm) and gradually add a PMMA cladding layer (n = 1.49) over that rod, increasing the coverage height H from 0 to 110 nm. This dielectric coating selectively enhances the local permittivity around the short rod, effectively increasing its effective dipole moment $p_1$. As $p_1$ approaches $p_2$ in magnitude with out of phase, the effective total dipole moment of the system approaches zero once again, restoring the destructive interference condition. As a result, the system transitions to a new restored symmetry-protected BIC (RSP-BIC) state, and the resonance disappears in the spectrum, as shown in the bottom panel of Figure 5a.

To quantify this transition, we extracted the Q-factors from the spectra across a range of $L_1$ and H values. Figure 5c shows the Q-factor map, clearly indicating the region where a RSP-BIC emerges. At this RSP-BIC condition, the Q-factor exceeds $10^7$, substantially higher than surrounding regions, confirming that a permittivity-induced compensation can effectively reestablish the BIC condition from a geometrically asymmetric structure.

Crucially, unlike conventional BIC states, which are isolated from the external environment and remain intrinsically stable, the RSP-BIC is environmentally accessible. Small changes in the surrounding refractive index can significantly perturb this balance, driving a transition from the RSP-BIC back to a qBIC state. This approach, which leverages environmental RI to reversibly tune the system between RSP-BIC and qBIC states, offers a novel mechanism that has not been addressed in prior studies and may open new directions in metasurface-based sensing.

## 3. Conclusion

In this work, a permittivity-driven quasi-bound state in the continuum metasurface design is presented, where the environmental refractive index is directly encoded into the system's asymmetry factor, enabling not only the resonance wavelength but also the resonance intensity to vary systematically with the surrounding medium. Under a single-wavelength condition, the intensity variation response with fine RI steps are analyzed. Although both ε-qBIC and g-qBIC exhibit high sensitivity, ε-qBIC demonstrates a considerably broader and more stable linear window across the spectrum, suggesting enhanced robustness and reliability. Importantly, numerical results reveal that permittivity asymmetry can be used not only to generate a quasi-BIC from a symmetric BIC, but also to optically restore a geometrically symmetry broken qBIC into a new BIC state with $Q > 10^7$ that remains accessible to environmental changes. Unlike conventional BICs, this restored BIC is responsive to RI variations and represents an optical state between nonradiative and radiative regimes, broadening the design space for BIC-based photonics, offering a versatile framework for high-Q devices that merge fundamental control over light–matter interaction with practical sensing capabilities

## 4. Method

### 4.1 Optical Characterizations

Transmittance spectra were measured using a white-light transmission microscope (Witec Alpha Series 300). The samples were illuminated with linearly polarized white light, and the transmitted signal was collected using a 20X immersion objective with a numerical aperture of 0.5. The collected light was coupled into a multimode fiber and directed to a grating-based spectrometer equipped with a silicon CCD detector. All measurements were normalized to its corresponding signal from a bare fused silica substrate. For the initial sensing experiments in Figure 2, commercial refractive index liquids from Cargille Labs (Series: AAA) were used, with refractive indices ranging from n = 1.30 to 1.39 in steps of 0.01. To further evaluate the sensing performance in Figure 3 and 4, additional refractive index liquids were prepared by mixing water and glycerol to obtain a finer range of indices between n = 1.386 and 1.414, with an increment of $\Delta n = 0.004$.

### 4.2 Numerical simulations

Simulations were performed using CST Studio Suite, a commercial finite-element-based solver. The model was configured in the frequency domain with periodic boundary conditions and employed adaptive mesh refinement to ensure convergence. The dielectric functions of $TiO_2$ and PMMA used in the simulations were obtained from spectroscopic ellipsometry measurements and subsequently imported into CST. For sensing simulations, the background refractive index was varied to simulate changes in the surrounding environment.

### 4.3 Nano fabrications

Fused silica substrates were first cleaned in an ultrasonic acetone bath, then rinsed with isopropanol (IPA) and treated with oxygen plasma to remove any remaining contaminants. A 110 nm $TiO_2$ layer was then deposited by sputtering a titanium target in an oxygen-containing plasma (Angstrom Engineering). A layer of PMMA 950k A4 resist was spin-coated and baked at 180 °C for three minutes. To avoid charging during electron beam lithography (EBL), a conductive polymer (E-Spacer 300Z) was spin-coated on top of the PMMA. In the initial patterning step, a 30 nm gold markers system was defined on the $TiO_2$ film for alignment in the following processes. Using these markers, the second patterning step positioned two-rod nanostructure metasurfaces. After exposure, the sample was developed for 135 seconds in a 3:1 IPA–MIBK solution. A 50 nm chromium layer was then deposited by e-beam evaporation to serve as a hard mask, and lift-off took place in Microposit Remover 1165 at 80 °C overnight. The structures were transferred into the $TiO_2$ layer by reactive ion etching (RIE) in a PlasmaPro 100 ICP-RIE (Oxford Instruments), and the chromium mask was then removed in a wet Cr etchant. In the final patterning step, another PMMA layer was spin-coated and patterned onto the fabricated BIC metasurfaces using the same marker system. The patterned regions were cleaned using the same development protocol.

**Research funding**
This project was funded by the Deutsche Forschungsgemeinschaft (German Research Foundation) under grant numbers EXC 2089/1–390776260 (Germany's Excellence Strategy) and TI 1063/1 (Emmy Noether Program), the Bavarian programme Solar Energies Go Hybrid (SolTech), and the Center for NanoScience at LMU. It was also funded by the European Union (ERC, METANEXT, 101078018, and EIC, OMICSENS, 101129734) and Chinese scholarship council. Views and opinions expressed are however those of the author(s) only and do not necessarily reflect those of the European Union, the European Research Council Executive Agency, or the European Innovation Council and SMEs Executive Agency (EISMEA). Neither the European Union nor the granting authority can be held responsible for them.

**Author contributions**
H.H., A.T., X.Y., and A.A. contributed to the conceptual design of the study. X.Y. carried out the simulations, experiments, and data analysis, with support from H.H. and A.A. X.Y. and H.H. prepared the initial draft of the manuscript. All authors contributed to the methodology, critically revised the manuscript, and approved the final version. A.T. supervised the project. All authors take full responsibility for the content of the paper.

**Acknowledgements**
The authors thank Jonas Biechteler and Maxim Gorkunov for valuable discussions.

**Conflict of interests**
Authors declare that they have no competing interests.

**Data availability**
The data that support the findings of this study are available from Zenodo [Zenodo link to be added at acceptance].

Permittivity-asymmetric qBIC metasurfaces for refractive index sensing

Supplementary information

Tables of contents

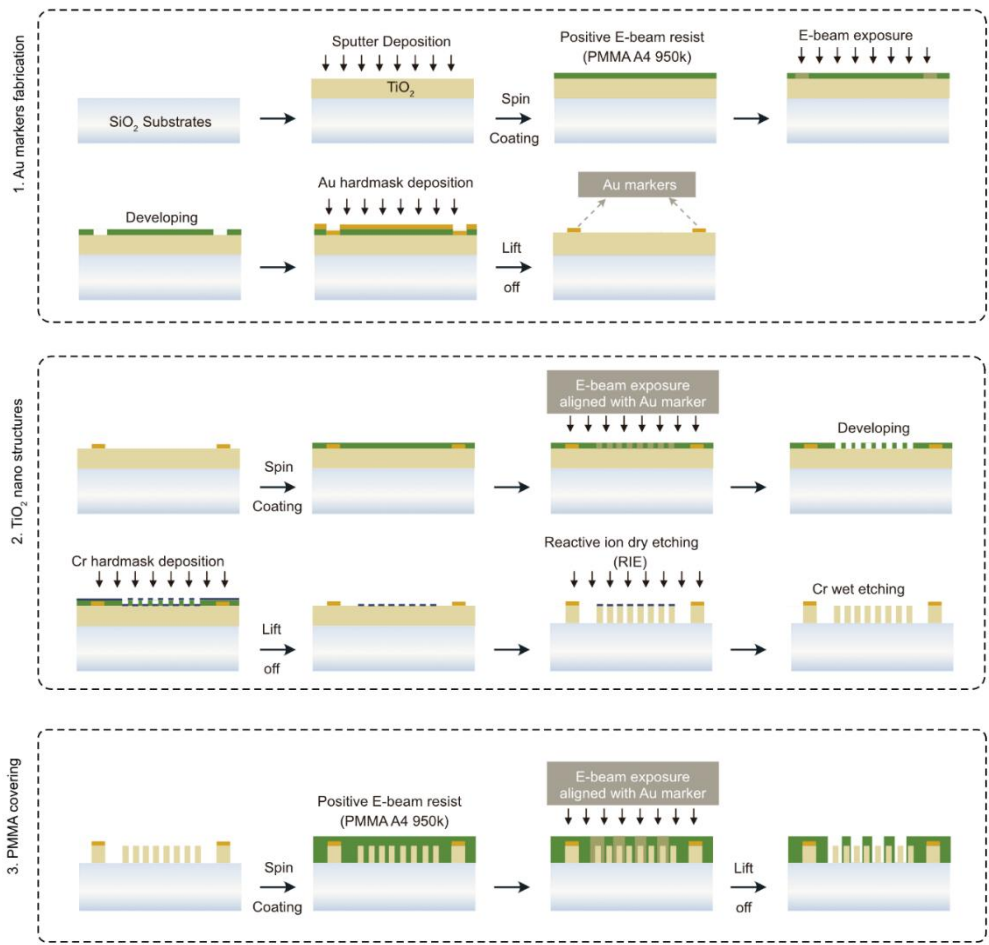

**Figure S1. Schematic overview of nanofabrication for ε-*q*BIC metasurface.** It includes three main steps, the fabrication of Au marker systems, TiO$_2$ nanostructures, and selective patterning on PMMA.

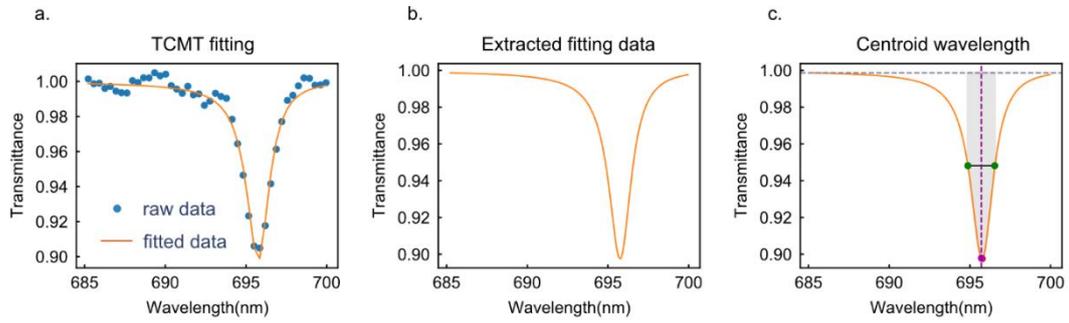

**Figure S2. Centroid wavelength calculation.** a. Raw discrete transmittance data fitted using the temporal coupling model theory (TCMT) to fit for the raw discrete spectral data. b. Clean fitted curve extracted from the TCMT model, applied for subsequent analysis. c. Illustration of centroid wavelength extraction from the fitted spectrum. The shaded gray area represents the integration region, and the vertical dashed line marks the calculated center of mass. This method follows established procedures reported in prior studies.

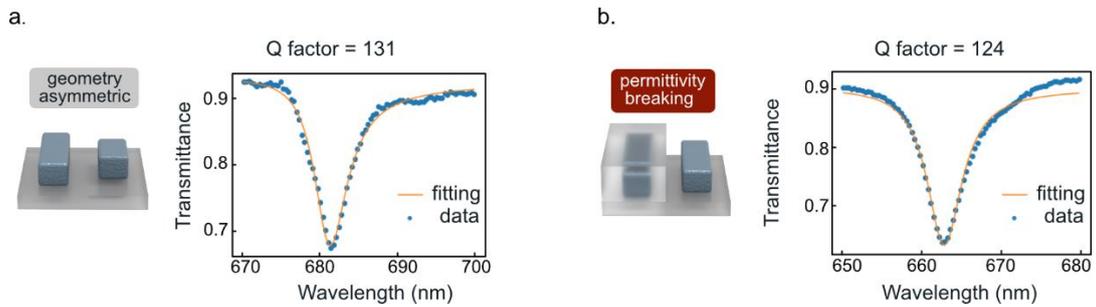

**Figure S3. Measured transmittance spectra of metasurfaces in the air environment.** a. The transmittance spectra of g-$q$BIC metasurface with Q factor 131. b. The transmittance spectra of ε-$q$BIC metasurface with Q factor of 124.

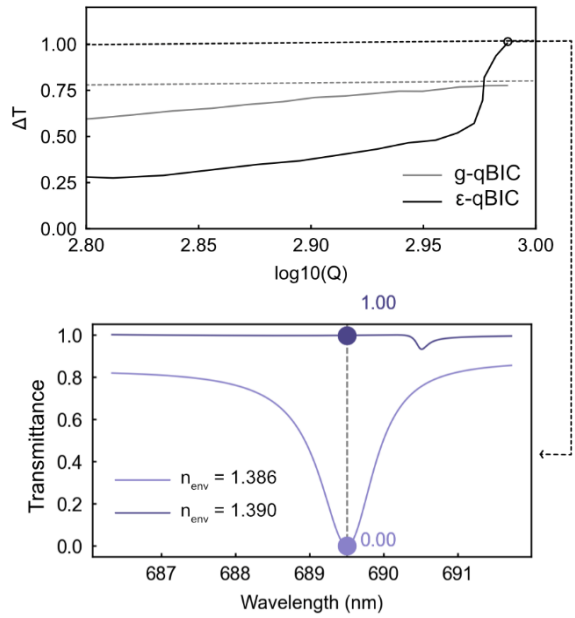

**Figure S4. Simulation of Q-factor vs intensity modulation response for g-qBIC and ε-qBIC.**

Simulations were performed to analyze the sensitivity performance of g-qBIC and ε-qBIC structures by sweeping the geometric asymmetry in g-qBIC and the cladding refractive index ($n_{cov}$) in ε-qBIC. For each configuration, the Q-factor was extracted from the simulated transmittance spectra. Subsequently, the environmental refractive index was varied from 1.386 to 1.390 ($\Delta n = 0.004$) to evaluate the corresponding intensity modulation signal $\Delta T$ at a fixed wavelength. The results show that in ε-qBIC, when $n_{cov}$ is tuned to closely match the surrounding environment, the system enters a high-Q regime where even a small RI change can strongly affect the radiative loss channel. When $n_{cov}$ equals the environmental index, the qBIC resonance can be even restored into a BIC state. This leads to an abrupt transition from a resonant to a non-resonant state, producing a unity modulation depth. In contrast, the $\Delta T$ signal in g-qBIC saturates at approximately 0.77.